\documentclass[a4paper,fleqn,usenatbib,useAMS]{mnras}                                  

\usepackage{graphicx}	
\usepackage{amsmath}	
\usepackage{amssymb}	
\usepackage{multicol}        
\usepackage{bm}		
\usepackage{pdflscape}	
\def\del#1{{}}

\newcommand{\dd}{\mathrm{d}}

\usepackage[T1]{fontenc}
\usepackage{ae,aecompl}

\title[A Phenomenological Model for the Intracluster Medium]{A Phenomenological Model for the Intracluster Medium that matches X-ray and Sunyaev-Zel'dovich observations}
\author[F. Zandanel, C. Pfrommer and F. Prada]{
Fabio Zandanel$^{1,2}$, Christoph Pfrommer$^{3}$ and Francisco Prada$^{4,5,1}$\\
$^{1}$Instituto de Astrof\'{\i}sica de Andaluc\'{\i}a (CSIC), Glorieta de la Astronom\'{\i}a, E-18080 Granada, Spain\\
$^{2}$Now at GRAPPA Institute, University of Amsterdam, Science Park 904, 1098XH Amsterdam, Netherlands, f.zandanel@uva.nl\\
$^{3}$Heidelberg Institute for Theoretical Studies, Schloss-Wolfsbrunnenweg 35, D-69118 Heidelberg, Germany, christoph.pfrommer@h-its.org\\
$^{4}$Campus of International Excellence UAM+CSIC, Cantoblanco, E-28049 Madrid, Spain\\
$^{5}$Instituto de F\'{\i}sica Te\'orica, (UAM/CSIC), Universidad Aut\'onoma de Madrid, Cantoblanco, E-28049 Madrid, Spain}
\begin{document}

\date{Accepted 2013 November 17}

\pagerange{\pageref{firstpage}--\pageref{lastpage}} \pubyear{2016}

\maketitle

\label{firstpage}

\begin{abstract}
    Cosmological hydrodynamical simulations of galaxy clusters are still
    challenged to produce a model for the intracluster medium that matches all
    aspects of current X-ray and Sunyaev-Zel'dovich observations. To facilitate
    such comparisons with future simulations and to enable realistic cluster
    population studies for modeling e.g., non-thermal emission processes, we
    construct a phenomenological model for the intracluster medium that is based
    on a representative sample of observed X-ray clusters. We create a mock
    galaxy cluster catalog based on the large collisionless $N$-body simulation
    MultiDark, by assigning our gas density model to each dark matter cluster
    halo.  Our clusters are classified as cool-core and non cool-core according
    to a dynamical disturbance parameter. We demonstrate that our gas model
    matches the various observed Sunyaev-Zel'dovich and X-ray scaling relations
    as well as the X-ray luminosity function, thus enabling to build a reliable
    mock catalog for present surveys and forecasts for future experiments. In a
    companion paper, we apply our catalogs to calculate non-thermal radio and
    gamma-ray emission of galaxy clusters. We make our cosmologically complete
    multi-frequency mock catalogs for the (non-)thermal cluster emission at
    different redshifts publicly and freely available online through the
    MultiDark database.
\end{abstract}


\section{Introduction}
\label{sec:1}

  Galaxy clusters are the rarest and largest gravitationally-collapsed
  objects in the Universe and form at sites of constructive interference of long
  waves in the primordial density fluctuations. Clusters reach radial extends of
  a few Mpc and total masses $M \sim (10^{14} - 10^{15}) M_{\odot}$, of which
  galaxies, hot ($1-10$~keV) gas and dark matter (DM) contribute roughly 5, 15
  and 80 per cent, respectively. If the thermal properties of the intracluster medium
  (ICM) were solely determined by the gravity of the system, then clusters would
  obey self-similar scaling relations \citep{1986MNRAS.222..323K}.  However,
  X-ray observations have demonstrated that self-similarity is broken,
  especially at low-mass scales (see, e.g., \citealp{2005RvMP...77..207V} for a
  review). Apparently the energy input from non-gravitational processes
  associated with galaxy formation have a larger influence in smaller systems,
  i.e., at scales of galaxy groups.

  The central cooling time of the X-ray emitting gas in galaxy clusters and
  groups is bimodally distributed. Approximately half of all systems have
  radiative cooling times of less than 1~Gyr and establish a population of cool
  core clusters (CCCs), while others have cooling times that can be longer than
  the age of the Universe, forming the non-cool core cluster (NCCC) population
  \citep{2009ApJS..182...12C,2010A&A...513A..37H}. Several physical processes
  have been proposed to be responsible for balancing radiative cooling of the
  low-entropy gas at the centers of CCCs. In particular feedback by active
  galactic nuclei (AGN) has come into the focus of recent research
  \citep{2007ARA&A..45..117M, 2012NJPh...14e5023M}, owing to the self-regulated
  nature of the proposed feedback and observational correlations between radio
  activity and central entropy (which is a proxy for a small cooling time). The
  underlying physical processes responsible for the heating include dissipation
  of mechanical heating by outflows, lobes, or sound waves from the central AGN
  \citep[e.g.,][]{2001ApJ...554..261C, 2002Natur.418..301B, 2002ApJ...581..223R,
    2012MNRAS.424..190G} or cosmic-ray heating: streaming cosmic rays excite
  Alfv\'en waves on which they resonantly scatter \citep{1969ApJ...156..445K}.
  Damping of these waves transfers cosmic ray energy and momentum to the cooling
  plasma \citep{1991ApJ...377..392L, 2008MNRAS.384..251G, 2011A&A...527A..99E,
    2013MNRAS.434.2209W}, heating it at a rate that balances radiative cooling
  on average at each radius while explaining the observed temperature floor in
  CCCs by thermal stability of the heating mechanism
  \citep{2013arXiv1303.5443P}.

  Modeling the formation and evolution of clusters with cosmological
  hydrodynamical simulations has been progressively refined over the last years
  to also include (sub-resolution models for) AGN feedback
  \citep[e.g.,][]{2007MNRAS.380..877S, 2008MNRAS.387.1403S, 2008ApJ...687L..53P,
    2012MNRAS.420.2662D, 2012MNRAS.424..190G, 2013MNRAS.428.2366V}. While the
  comparison between data and simulations with AGN feedback improved for integrated
  thermal properties of large cluster samples
  \citep{2012ApJ...758...74B,2012ApJ...758...75B}, there are still notable
  differences for differential quantities such as the pressure profile
  \citep{2013A&A...550A.131P} or the profile of gas mass fractions
  \citep{2012arXiv1209.4082B}. These discrepancies (that have been more dramatic
  in the past without AGN feedback) motivated the development of
  phenomenological models of the ICM \citep[e.g.,][]{2005ApJ...634..964O,
    2012MNRAS.422..686C}. Here, we construct a simple and purely
  phenomenological model that matches all available X-ray and Sunyaev-Zel'dovich
  (SZ) data with the goal to facilitate comparison with future hydrodynamical
  simulations, to allow the modeling of non-thermal cluster emission over the
  entire electromagnetic wave-band, and finally to enable the construction of
  cluster mock catalogs of the thermal and non-thermal cluster emission (in the
  radio, hard X-ray, and gamma-ray band) of current and future surveys. Such a
  model for the ICM needs to be applied to a sample of cluster halos, which
  can either be obtained by means of analytical expressions for the cluster mass
  function (e.g., \citealp{2001MNRAS.321..372J}) or cosmological
  simulations. In this work, we make use of the recent large-volume $N$-body
  simulation MultiDark, which employs a flat $\Lambda$ Cold Dark Matter cosmology
  \citep{2011arXiv1104.5130P}.

  Our \emph{phenomenological} approach uses gas density profiles of a
  representative sample of observed X-ray clusters, which are complemented by
  the observed cluster mass-dependent gas mass fractions. The resulting cluster
  mock sample are then compared to data including the X-ray luminosity function
  (XLF), the luminosity-mass relation, $L_{\rmn{X}}- M$, and the $Y_{\rmn{X}}-M$
  relation, where $Y_{\rmn{X}}=M_\rmn{gas}k_{\rmn{B}}T$ with an X-ray-derived
  gas mass $M_{\rmn{gas}}$ and temperature $T$ \citep{2006ApJ...650..128K}. We
  additionally compare our $Y_{\rmn{SZ}}-M$ relation to recent SZ data, where
  $Y_\rmn{SZ}$ denotes the integrated Compton-$y$ parameter.  In a companion
  paper (\citealp{paper2}; hereafter Paper~II), we apply our ICM model to
  predict the non-thermal radio and gamma-ray emission of a cosmological
  complete sample of galaxy clusters, enabling valuable insight into the
  statistics of radio halos.  One of our final data products is a complete
  cosmological cluster mock catalog, at different redshifts, containing
  information of the DM and ICM properties for each object, together with
  its (non-)thermal emission at different frequencies. All the products can
    be found on-line at the MultiDark database (\textit{www.multidark.org}).

    In Section~\ref{sec:2}, we introduce the MultiDark simulation along with our
    selected cluster halo sample. We assign to each of our clusters the
    phenomenologically constructed gas density profile in
    Section~\ref{sec:3}. In Sections~\ref{sec:4} and \ref{sec:5}, we show that
    this approach can successfully reproduce the observed X-ray and SZ cluster
    data. We present the resulting mock cluster catalogs in Section~\ref{sec:6},
    and eventually summarize in Section~\ref{sec:7}.  In this work, the cluster
    mass $M_{\Delta}$ and radius $R_{\Delta}$ are defined with respect to a
    density that is $\Delta=200$ or 500 times the \emph{critical} density of the
    Universe. We adopt density parameters of $\Omega_{\rmn{m}}=0.27$,
    $\Omega_{\Lambda}=0.73$ and today's Hubble constant of $H_0 = 100\,
    h_{70}$~km~s$^{-1}$~Mpc$^{-1}$ where $h_{70} = 0.7$.

\section{MultiDark simulation and final cluster sample}
\label{sec:2}
The MultiDark simulation used in this work is described in detail in
\cite{2011arXiv1104.5130P}.  It is an $N$-body cosmological simulation done with
the Adaptive-Refinement-Tree (ART) code \citep{1997ApJS..111...73K} of $2048^3$
particles within a ($1000\,h^{-1}$~Mpc)$^3$ cube. The adopted cosmological
  parameters are $\Omega_{\rmn{m}}=0.27$, $\Omega_{\Lambda}=0.73$,
  $\Omega_\rmn{b}=0.0469$, $n_\rmn{s}=0.95$, $h=0.7$ and $\sigma_8=0.82$. This
simulation is particularly well suited for our purpose because of its large
number of high-mass objects, i.e., galaxy clusters.
 
We use the MultiDark halo catalog from its on-line database,\footnote{\textit{www.multidark.org} \citep{2013AN....334..691R}} 
constructed with the Bound Density Maxima (BDM) algorithm \citep{1997astro.ph.12217K}.  
We will mainly use $M_{500}$ and $R_{500}$ for comparison with existing observational works.  
We use the technique described in \cite{2003ApJ...584..702H} to convert $M_{200}$ and
$R_{200}$ provided by the MultiDark halo catalog to $M_{500}$ and $R_{500}$.  In
creating our cluster sample we only select distinct halos, i.e., those halos that
are not sub-halos of any other halo, which by definition are not galaxy clusters.

Additionally, we assume that the main emission mechanism in the ICM is thermal
bremsstrahlung, which is true only above a particle energy of approximately
$2.6$~keV \citep{1988xrec.book.....S}. Below this
energy, there could be other important contributions to the emission,
e.g., from atomic lines. Therefore, we impose a mass cut of
$M_{200}\geq1\times10^{14}$~$h^{-1}$~M$_{\odot}\approx1.4\times10^{14}$~$h_{70}^{-1}$~M$_{\odot}$
which ensures $k_{\rmn{B}}T \gtrsim 2.6$~keV (assuming the $M_{500} - T_{\rmn{ci}}$ relation
of \citealt{2010MNRAS.406.1773M}).

  In Paper~II, we present predictions for the LOFAR radio observatory which
  expects to detect diffuse radio emission in clusters up to redshift $z \approx
  1$ (e.g., \citealp{2012JApA..tmp...34R}). Thus, we make use of different
simulation snapshots up to $z=1$.  The extrapolation of our model beyond
  this redshift is rather uncertain as it is based on observations at low(er)
  redshift. In Table~\ref{tab:z}, we show the total cluster number in our final
cluster sample at different redshifts.

\begin{table} 
\begin{center}
\caption{Number of halos in the final cluster sample}
\medskip
\begin{tabular}{cc}
\hline
\phantom{\Big|}
redshift $z$ & number of halos \\
\hline\\[-0.5em]
 0.0~~ &  13763\\
 0.1~~ &  12398\\
 0.18 &  11106\\
 0.2~~ &  10783\\ 
 0.4~~ &   ~~7789\\
 0.53 & ~~6079\\ 
 0.61 &  ~~5187\\ 
 0.78 &  ~~3372\\ 
 1.0~~ &  ~~1803\\[0.5em]
\hline
\end{tabular}
\label{tab:z}
\end{center}
\footnotesize{Note. We show the number of halos in our MultiDark snapshots at redshift $z$ for $M_{200}\geq1\times10^{14}$~$h^{-1}$~M$_{\odot}\approx1.4\times10^{14}$~$h_{70}^{-1}$~M$_{\odot}$.  More snapshots can be found online at \emph{www.mutlidark.org}.}
\end{table}

\section{Gas density modeling}
\label{sec:3}

\begin{figure} 
\centering
\includegraphics[width=0.48\textwidth]{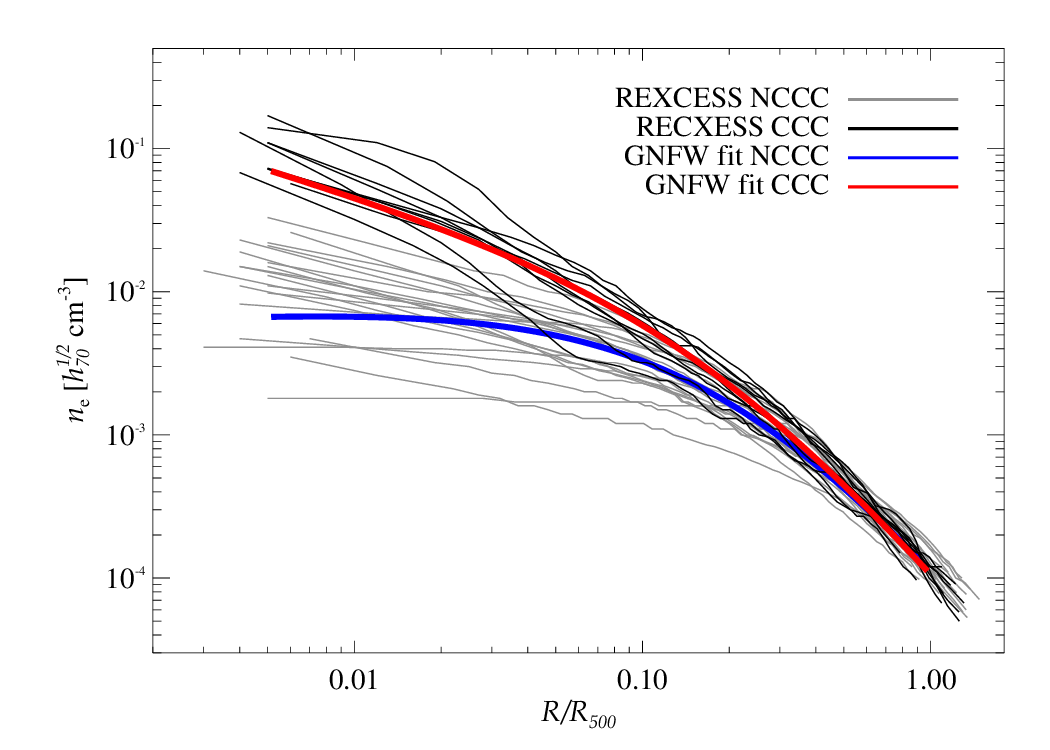}
\caption{Electron density profiles of the 31 clusters in the REXCESS sample. Grey and black lines represent 
NCCCs and CCCs, respectively. The blue and red lines represent our GNFW mean profile for the NCCCs and CCCs, respectively.}
\label{fig:gas_profiles}
\end{figure}

We decided to use a \emph{phenomenological} approach and construct the gas
density profiles directly from X-ray observations. A suitable X-ray sample that
provides the required information is the \emph{Representative XMM-Newton Cluster
  Structure Survey} (REXCESS) sample \citep{2008A&A...487..431C,
  2009A&A...498..361P}. It is a sample of 31 galaxy clusters of different
dynamical states at redshifts $0.06<z<0.18$ with detailed information on the
de-projected electron density profiles \citep{2008A&A...487..431C}. In
Fig.~\ref{fig:gas_profiles}, we show the 31 electron density profiles of the
REXCESS sample color-coded by NCCCs and CCCs.

In order to obtain an electron density profile that we will attach to our
simulated clusters, we use a generalized Navarro-Frank-White (GNFW) profile,
\begin{equation}
n_{\rmn{e}}(x) = \frac{n_{0}}{x^{\beta}\left[1+x^{\alpha}\right]^{\frac{\delta-\beta}{\alpha}}},
\label{eq:gnfw}
\end{equation}
where $x=R/R_{\rmn{c}}$ and $R_{\rmn{c}}$ is the cluster core radius. To reduce
the dimensionality of our fit, we fix representative values of $R_{\rmn{c}} =
0.2\, R_{500}$, $\alpha = 1$ and $\delta = 2.5$. We fit the radial density
profiles of the REXCESS sample in log-log space, separating them in the two
categories of NCCCs and CCCs as shown in Fig.~\ref{fig:gas_profiles}.  We obtain
$n_{\rmn{0,NCCC}} = 1.02\times10^{-2}$~$h_{70}^{1/2}$~cm$^{-3}$,
$n_{\rmn{0,CCC}} = 8.32\times10^{-3}$~$h_{70}^{1/2}$~cm$^{-3}$,
$\beta_{\rmn{NCCC}} = -0.093$, and $\beta_{\rmn{CCC}} = 0.592$. The resulting
fits are shown in blue and red for the NCCC and CCC population,
respectively.

The next step is to introduce a mass-scaling in order to apply our GNFW profiles
to clusters of all masses. We adopt the gas mass fraction-mass scaling,
$f_{\rmn{gas},500}-M_{500}$ of \cite{2009ApJ...693.1142S} (and adopt their
equation~(8)). We can express $f_{\rmn{gas},500}$ in the following way:
\begin{equation}
f_{\rmn{gas},500} = \frac{M_{\rmn{gas},500}}{M_{500}}  = \frac{\int_{0}^{R_{500}} \rho_{\rmn{gas}} \rmn{d}V}{M_{500}}
\label{eq:m500}
\end{equation}
with $\rho_{\rmn{gas}}(R) = \rho_{\rmn{gas}} = n_{\rmn{e}} m_{\rmn{p}} / ( X_{\rmn{H}}X_{\rmn{e}} )$ where
$m_{\rmn{p}}$ is the proton mass, $X_{\rmn{H}} = 0.76$ is the primordial hydrogen
mass fraction and $X_{\rmn{e}} = 1.157$ the ratio of electron-to-hydrogen number
densities in the fully ionized ICM \citep{1988xrec.book.....S}. For each cluster
$i$ of our sample, we then define a \emph{mass-scaled} gas profile as
$\rho_{\rmn{gas},i}=C_{i} \,\rho_{\rmn{gas}}$ with:
\begin{eqnarray}
C_{i}  & = &  (0.0656\pm(0.0064g_{1})) \, h_{70}^{-1.5}  \\
 & \times & \left(\frac{M_{500,i}}{1.04 \times 10^{13} h_{70}^{-1} \rmn{M_{\odot}}}\right)^{0.135\pm(0.030g_{2})} \frac{M_{500,i}}{\int_{0}^{R_{500,i}} \rho_{\rmn{gas}} \rmn{d}V} \nonumber
\label{eq:gas_scaling}
\end{eqnarray}
where $g_{1}$ and $g_{2}$ are random Gaussian numbers, which we use in order to
simulate the natural scatter of the gas profiles.\footnote{The values
  $0.0064$ and $0.03$ quoted in equation~(\ref{eq:gas_scaling}) do not represent
  the proper scatter of the $f_{\rmn{gas},500}-M_{500}$ relation but reflect the
  parameter errors and we rescaled the numerical values to a Hubble constant of
  $h_{70}$ used in this work.}

Hence, for each cluster in our sample we obtain a gas density profile
$\rho_{\rmn{gas},i}$ that obeys the observed $f_{\rmn{gas},500}-M_{500}$ relation and
is uniquely determined by its total mass $M_{500,i}$ and by the property of being a
NCCC or CCC. We assign the latter property to every halo depending on its
merging history. In particular, we make use of the offset parameter
$X_{\rmn{off}}$ computed for the MultiDark halo catalog. This is defined as the
distance from the halo center to the center of mass in units of the virial
radius. This parameter assesses the dynamical state of the cluster and whether
the halo has experienced a recent merger or not. Current observations reveal a ratio
of NCCCs and CCCs of about 50 per cent (e.g., \citealp{2007A&A...466..805C,
  2009MNRAS.395..764S}). Since there is a correlation between merging clusters
and NCCCs, we use the median of the $X_{\rmn{off}}$ distribution to separate our
sample into CCCs and NCCCs (with NCCCs defined to be those halos with the larger
dynamical offsets). Clearly, this is an over-simplification, and future X-ray
surveys will have to determine this property also as a function of redshift.

\begin{figure*} 
\centering
\includegraphics[width=0.48\textwidth]{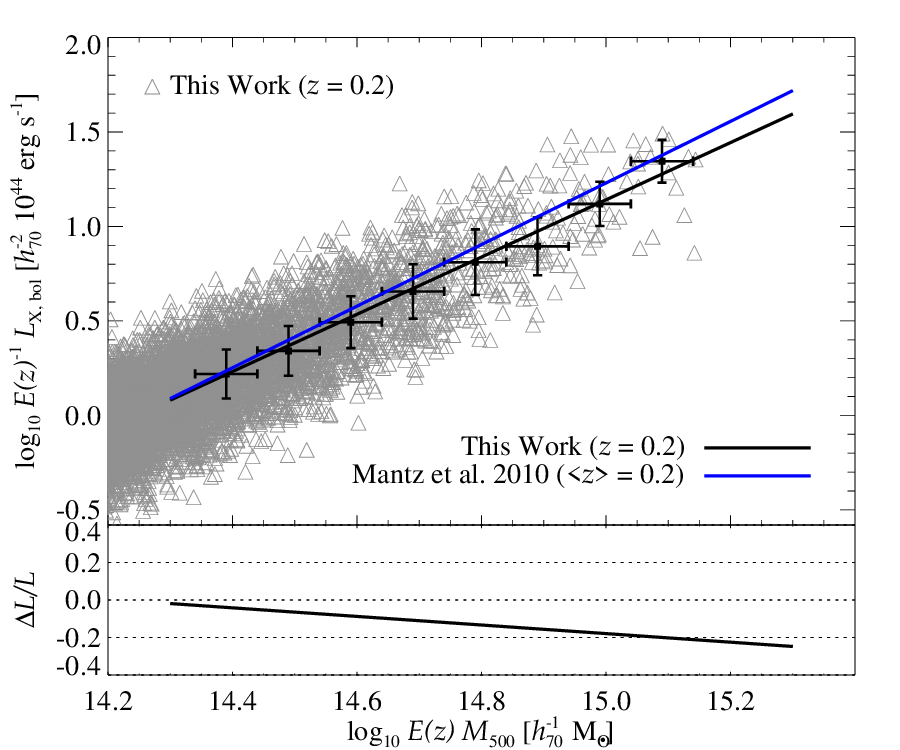}
\includegraphics[width=0.48\textwidth]{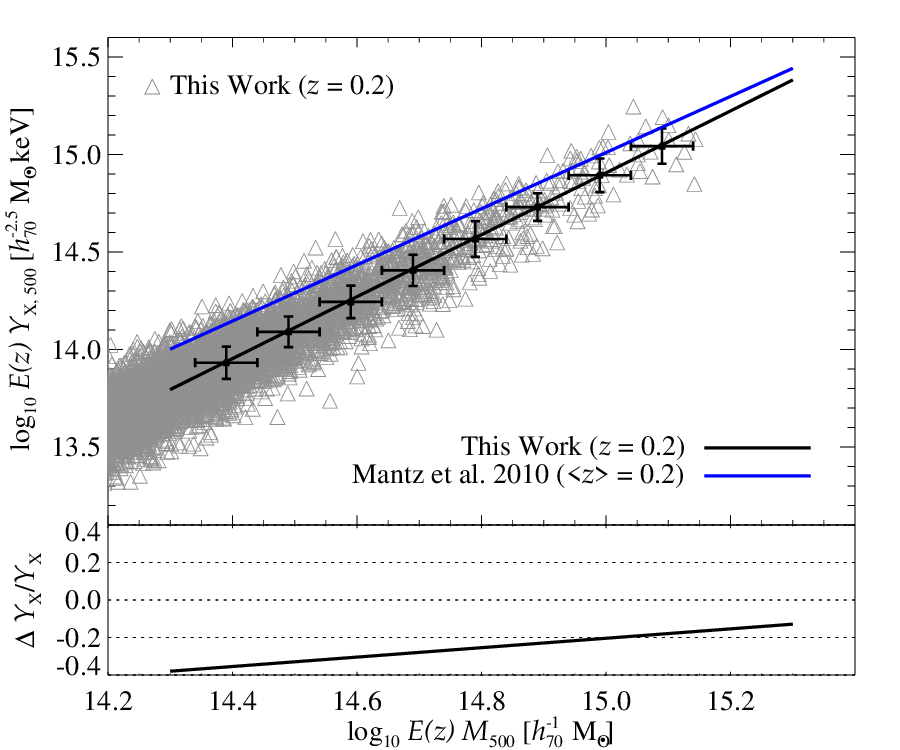}
\caption{X-ray cluster scaling relations. Grey triangles show the MultiDark
  sample (limited to the mass range covered by observations), the black line is
  the corresponding scaling relation, and the blue line is the
  observational result by \protect\cite{2010MNRAS.406.1773M}. 
  The black crosses represent the median values of the
  quantity in question for a given mass bin (indicated by horizontal error
  bars), and the vertical error bars represent the standard deviation within a
  bin. The bottom panels show the relative difference to the observational
    scaling relations. \emph{Left.} We compare the bolometric X-ray luminosity-to-mass
  relation, $L_{\rmn{bol}}-M_{500}$, at $z=0.2$ to the observational sample by
  \protect\cite{2010MNRAS.406.1773M} with a median of $z \approx 0.2$. 
  \emph{Right.} $Y_{\rmn{X}}-M_{500}$ scaling relation of our model in comparison to the
  observational sample by \protect\cite{2010MNRAS.406.1773M}.} 
\label{fig:X_LM}
\end{figure*}

We also account for redshift evolution of the gas profiles. While our NCCC and
CCC gas profiles as derived from the REXCESS cluster sample are merely used to
define a profile shape, the normalization of the gas profiles is set by the
observational $f_{\rmn{gas},500}-M_{500}$ relation \citep{2009ApJ...693.1142S}.
The 43 clusters used in \cite{2009ApJ...693.1142S} have redshifts $0.012 < z <
0.12$ with a \emph{median} of $z \approx 0.04$. Thus, our phenomenological gas
profile is representative of the cluster population at $z\approx0$. To extend
this profile to high-$z$, we adopt a \emph{self-similar} scaling of the gas
density as $\rho_{\rmn{gas}}(z) = E(z)^{2} \rho_{\rmn{gas}}(z=0)$, where
$E(z)^{2} = \Omega_{\rmn{m}} (1+z)^{3} + \Omega_{{\Lambda}}$.

  As a cautionary remark, we emphasize that our model has been constructed to
  be valid within $R_{500}$ and more work is needed to extrapolate it to larger
  cluster radii, particularly for the thermal cluster emission.  Those cluster
  regions are characterized by a steadily increasing level of kinetic-to-thermal
  pressure support \citep[e.g.,][]{2009ApJ...705.1129L, 2012ApJ...758...74B},
  density clumping \citep{2011ApJ...731L..10N, 2012arXiv1209.4082B}, and
  asphericity and ellipticity of the halo morphologies
  \citep{2012arXiv1209.4082B}, in particular beyond the virial radius where the
  filamentary cosmic web connects to the cluster interiors.

\section{X-ray  observables}
\label{sec:4}

So far, we used a well-observed representative sample of X-ray clusters
(REXCESS) that was supplemented with X-ray data on gas mass fractions to construct
our phenomenological gas model.  Here, we will test how our cluster mock data
compare to various X-ray inferred observables, which are obtained from different
cluster samples and from (partially) different X-ray observatories. Those
include the observed $L_{\rmn{bol}} - M_{500}$ relation, the $Y_X - M_{500}$
relation, and the XLF.

\subsection{Scaling Relations}
\label{sec:Xscaling}

First, we calculate the bolometric thermal bremsstrahlung luminosity
$L_{\rmn{bol}}$ following \cite{1988xrec.book.....S}.\footnote{We check our
  procedure by fitting each of the 31 REXCESS clusters with
  equation~(\ref{eq:gnfw}) and calculate $L_{\rmn{bol}}$ using the measured gas
  temperature of each cluster. As a result, we fall short of the observed
  luminosity by a mean (median) of about 21 per cent (20 per cent). This is
  acceptable considering that we do not permit the parameters $R_{\rmn{c}}$,
  $\alpha$ and $\delta$ to vary between different objects. Additionally, we
  neglect atomic line emission which may give a noticeable contribution, in
  particularly for low-mass clusters and in the cluster outskirts of larger
  systems.}  To assign a temperature to our model clusters (that is needed for
calculating $L_{\rmn{bol}}$ and $Y_{\rmn{X}}$), we adopt the $T-M_{500}$
relation by \cite{2010MNRAS.406.1773M},
\begin{equation}
\log_{10} \left( \frac{k_{\rmn{B}}T_{\rmn{ci}}}{\rmn{keV}} \right) = 
A + B~\log_{10} \left( \frac{E(z) M_{500}}{10^{15} h_{70}^{-1} \rmn{M_{\odot}}} \right)
\label{eq:temp}
\end{equation}
where $A=0.91$, $B=0.46$, and $T_{\rmn{ci}}$ is the cluster temperature
\emph{not} centrally excised
(\citealp{2010MNRAS.406.1773M}). \cite{2010MNRAS.406.1773M} report a scatter of
$\sigma_{\rmn{yx}} = 0.06,$\footnote{Scatter is calculated as $\sigma_{\rmn{yx}}
  = \sqrt{ \left\{ \Sigma_{i=1}^{N} [Y_{i}-(A+B~X_{i})]^{2}\right\} / N-1}$
  where the sum extends over the data points $X_{i}, Y_{i}$, and $A$ and $B$ are
  the fit parameters.} which we apply to our sample using Gaussian deviates.

In the left panel of Fig.~\ref{fig:X_LM}, we show how our model
$L_{\rmn{bol}}-M_{500}$ relation compares with observations by
\cite{2010MNRAS.406.1773M} (\emph{all} data, see their Table~7). Their sample is
composed of 238 clusters at $0.02<z<0.46$ with a median of $z \approx 0.2$ and
self-consistently takes into account all selection effects, covariances,
systematic uncertainties and the cluster mass function
\citep{2010MNRAS.406.1759M}.  For this reason, we compare the
\cite{2010MNRAS.406.1773M} data to our model at $z=0.2$, and limit the
comparison to the mass range covered by the observations. Overall, there is
reassuring agreement between our phenomenological model and the data, which
probe our model most closely on scales around the cluster core radii (which is
where the contribution to $L_{\rmn{X}}$ per logarithmic interval in radius, $\dd
L_{\rmn{X}}/\dd\log r \propto r^3 n_{\rm{gas}}^2(r) \sqrt{k_{\rmn{B}}T}$,
approximately attains its maximum).  In Table~\ref{tab:LMfits}, we show our
model $L_{\rmn{bol}}-M_{500}$ scaling relation and its scatter for different
redshifts. We find that the scatter of our samples at all redshifts are Gaussian
distributed with a standard deviation of $\sigma_{yx} \approx 0.18$ that matches
the observational results of \cite{2010MNRAS.406.1773M}, which report a scatter
of $\sigma_{yx} = 0.185$.

In the right panel of Fig.~\ref{fig:X_LM}, we compare the $Y_{\rmn{X}}-M_{500}$
relation of our sample to observational data \citep{2010MNRAS.406.1773M}. The
model agrees nicely at the high-mass end, but underpredicts the observed scaling
at low masses. 
The differential contribution to the thermal energy per logarithmic interval in
radius (and hence to the integrated Compton-$y$ parameter) is given by $\dd Y
/\dd\log r \propto r^3 P_{\rmn{th}}(r)$, with the thermal gas pressure
$P_{\rmn{th}}=n_{\rmn{gas}}k_{\rmn{B}}T$. It peaks at scales slightly smaller
than $R_{500}$ with 1-$\sigma$ contributions extending out to $3\,R_{500}$
\citep{2010ApJ...725...91B}. Hence, the observational scaling constrains our
model on those large scales, quite complementary to the X-ray luminosity. The
deviations at small masses either indicate tensions with our assumptions on
$f_{\rmn{gas}}$, the gas temperature, or different selection effects of either
observational sample that we use for model calibration and comparison.
\cite{2010MNRAS.406.1773M} determine their masses by adopting a constant value
for $f_{\rmn{gas}}$, in contrast to our approach which adopts the observed
$f_{\rmn{gas},500}-M_{500}$ relation given by
\cite{2009ApJ...693.1142S}. Additionally, we adopt the
\cite{2010MNRAS.406.1773M} \emph{centrally included} temperature throughout all
our work, while \cite{2010MNRAS.406.1773M} use the \emph{centrally excised}
temperature to calculate $Y_{\rmn{X}}$. This assumption also impacts the scatter
of the $Y_{\rmn{X}}-M$ relation. In fact, using the \emph{centrally included}
temperature, we found a scatter of $\sigma_{yx} \approx 0.11$ (see
Table~\ref{tab:YXfits} where our $Y_{\rmn{X}}$ scaling relations are reported),
significantly higher than the value of $\sigma_{yx} = 0.052$ found by
\cite{2010MNRAS.406.1773M}.

\begin{table} 
\begin{center}
\caption{$L_{\rmn{bol}}-M_{500}$ scaling relations.}
\medskip
\begin{tabular}{cccc}
\hline
\phantom{\Big|}
redshift $z$ & $A$ & $B$ & $\sigma_{yx}$ \\
\hline \\[-0.5em]
 0      & $-21.41\pm0.11$ & $1.50\pm0.01$ & 0.179\\
 0.1   & $-21.33\pm0.12$ & $1.50\pm0.01$ & 0.179\\
 0.18   & $-21.56\pm0.13$ & $1.51\pm0.01$ & 0.177\\
 0.2   & $-21.58\pm0.13$ & $1.51\pm0.01$ & 0.178\\ 
 0.4   & $-21.30\pm0.17$ & $1.49\pm0.01$ & 0.178\\ 
 0.53   & $-21.68\pm0.20$ & $1.52\pm0.01$ & 0.175\\
 0.61 & $-21.87\pm0.22$ & $1.53\pm0.01$ & 0.177\\ 
 0.78 & $-21.08\pm0.29$ & $1.48\pm0.02$ & 0.177\\ 
 1      & $-20.91\pm0.42$ & $1.46\pm0.03$ & 0.177\\[0.5em]
\hline
\end{tabular}
\label{tab:LMfits}
\end{center}
\footnotesize{Note. Scaling relations are reported in the form of $\log_{10}~(L_{\rmn{bol}}~/~E(z)~h_{70}^{-2}~10^{44}~\rmn{erg~s}^{-1})=A+B~\log_{10}~(E(z)~M_{500}~/~h_{70}^{-1}~\rmn{M_{\odot}})$. The relation scatter $\sigma_{yx}$ is also shown.}
\end{table}
 
\begin{table} 
\begin{center}
\caption{$Y_{\rmn{X}, 500}-M_{500}$ scaling relations.}
\medskip
\begin{tabular}{cccc}
\hline
\phantom{\Big|}
redshift $z$ & $A$ & $B$ & $\sigma_{yx}$ \\
\hline\\[-0.5em]
 0      & $-9.09\pm0.07$ & $1.60\pm0.01$ & 0.109\\
 0.1   & $-8.94\pm0.07$ & $1.59\pm0.01$ & 0.109\\
 0.18   & $-8.97\pm0.08$ & $1.59\pm0.01$ & 0.109\\
 0.2   & $-8.91\pm0.08$ & $1.59\pm0.01$ & 0.109\\ 
 0.4   & $-8.94\pm0.10$ & $1.59\pm0.01$ & 0.108\\ 
 0.53   & $-9.05\pm0.12$ & $1.60\pm0.01$ & 0.108\\ 
 0.61 & $-9.01\pm0.14$ & $1.59\pm0.01$ & 0.109\\ 
 0.78 & $-8.76\pm0.18$ & $1.58\pm0.01$ & 0.109\\ 
 1      & $-8.76\pm0.26$ & $1.58\pm0.02$ & 0.109\\[0.5em]  
\end{tabular}
\label{tab:YXfits}
\end{center}
\footnotesize{Note. Scaling relations are reported in the form of $\log_{10}~(E(z)~Y_{\rmn{X},500}~/~h_{70}^{-2.5}~\rmn{M_{\odot}}~\rmn{keV})=A+B~\log_{10}~(E(z)~M_{500}~/~h_{70}^{-1}~\rmn{M_{\odot}})$. The relation scatter $\sigma_{yx}$ is also shown.}
\end{table}

\subsection{Luminosity Functions}
Studies of the XLF got out of fashion during the last years due to the
difficulties of using the X-ray luminosity for cosmological purposes. The X-ray
emissivity scales with the square of the gas density, which makes it subject to
density variations and clumping. This implies large scatter that causes a
large Malmquist bias and underlines the necessity of careful mock surveys that
need to address all systematics.

\begin{figure} 
\centering
\includegraphics[width=0.48\textwidth]{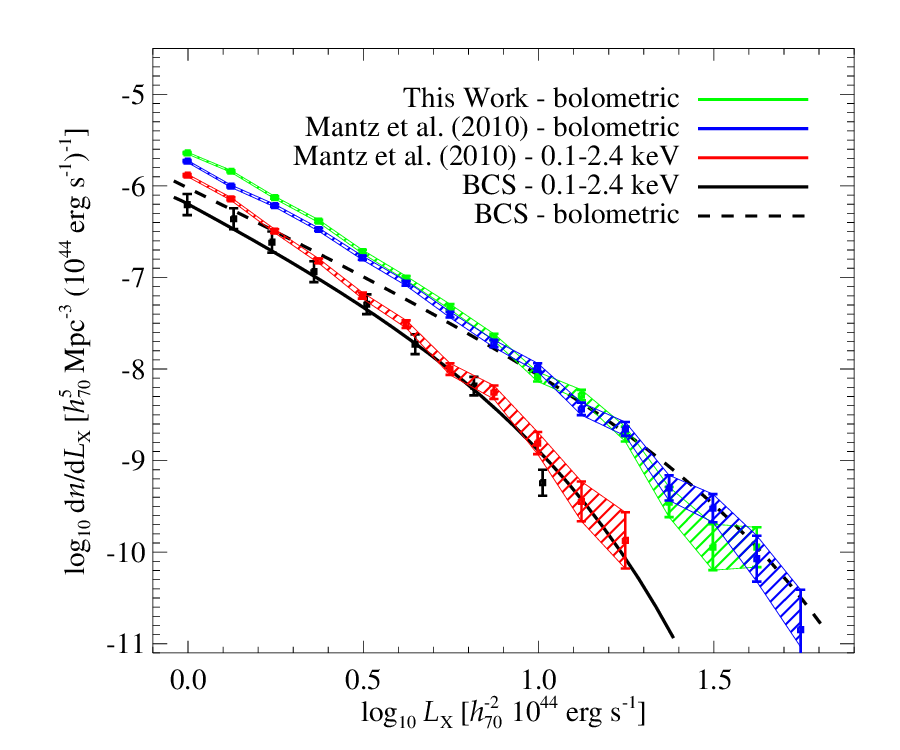}
\caption{Bolometric and soft-band ($0.1-2.4$~keV) XLFs. Shown are the soft-band
  data points, the soft-band and bolometric Schechter fits of the BCS sample of
  \protect\cite{1997ApJ...479L.101E}, which has a median of $z \approx 0.08$.
  While the soft-band XLF, which was obtained applying the
  \protect\cite{2010MNRAS.406.1773M} $L_{\rmn{X}}-M_{500}$ relation to the
  MultiDark $z = 0.1$ snapshot, compares well with the BCS data points, it
  deviates from the corresponding Schechter fit. We also show the bolometric XLF
  of \protect\cite{2010MNRAS.406.1773M} and the bolometric XLF of our model at
  $z=0.1$. The XLFs are calculated in equally log-spaced mass bins; the error
  bars represent the Poissonian errors. Note that we limit the comparison to the
  luminosity range covered by our sample.  We impose a low-luminosity cut to
  avoid a drop in the XLF due to the imposed mass cut.}
\label{fig:XLF}
\end{figure}

Nevertheless, it provides a complementary check for our model. To this end, we
use the XLF of the \emph{ROSAT} brightest cluster sample
\citep[BCS][]{1997ApJ...479L.101E}, which is in good agreement with results from
the \emph{ROSAT} ESO Flux-Limited X-ray (REFLEX; \citealp{2002ApJ...566...93B})
and HIFLUGCS \citep{2002ApJ...567..716R}.  Note that the XLF is fully determined
by the mass function and the $L_{\rmn{X}}-M_{500}$ relation after taking into
account the observational biases.\footnote{The mean (median) difference at $z=0$
  between $L_{\rmn{bol}}$ within $R_{200}$ or within $R_{500}$ is $\approx 5$
  per cent ($\approx 7$ per cent). While $L_{\rmn{bol}}$ refers to the quantity
  calculated within $R_{500}$, we note that the XLF for luminosities calculated
  within $R_{200}$ will be barely changed.} This means that applying the
Malmquist and Eddington-bias-corrected $L_{\rmn{X}}-M_{500}$ relation by
\cite{2010MNRAS.406.1773M} directly to the MultiDark mass function and
accounting for the observational scatter in $L_{\rmn{X}}-M_{500}$ should yield
an unbiased XLF.

We show the resulting bolometric and soft-band ($0.1-2.4$~keV) XLF in
Fig.~\ref{fig:XLF} and compare those to the corresponding BCS XLFs and to our
model predictions. Note that there is only the Schechter fit available for the
BCS bolometric XLF.  While the soft-band XLF by \cite{2010MNRAS.406.1773M}
agrees well with the BCS data points, it deviates from the corresponding
Schechter fit at low luminosities. This is also true in the bolometric band,
where the XLFs of \cite{2010MNRAS.406.1773M} and our model agree well, but
deviate from the BCS Schechter fit at low luminosities. This may be an artifact
due to the use of Schechter fit instead of the data points or may point to
incompleteness of the BCS sample. Note that the Poissonian errors of the XLF
obtained from the MultiDark simulation are a lower limit as we are neglecting
the uncertainty due to cosmic variance.  Studies of the XLF will become again an
important topic with the upcoming launch of the \emph{e}ROSITA satellite (e.g.,
\citealp{2011MSAIS..17..159C}) and further studies in this direction are
desirable. For these reasons, we do not show XLF predictions at other redshifts,
leaving this for a future study.

\section{SZ scaling relations}
\label{sec:5}

\begin{figure*} 
\centering
\includegraphics[width=0.48\textwidth]{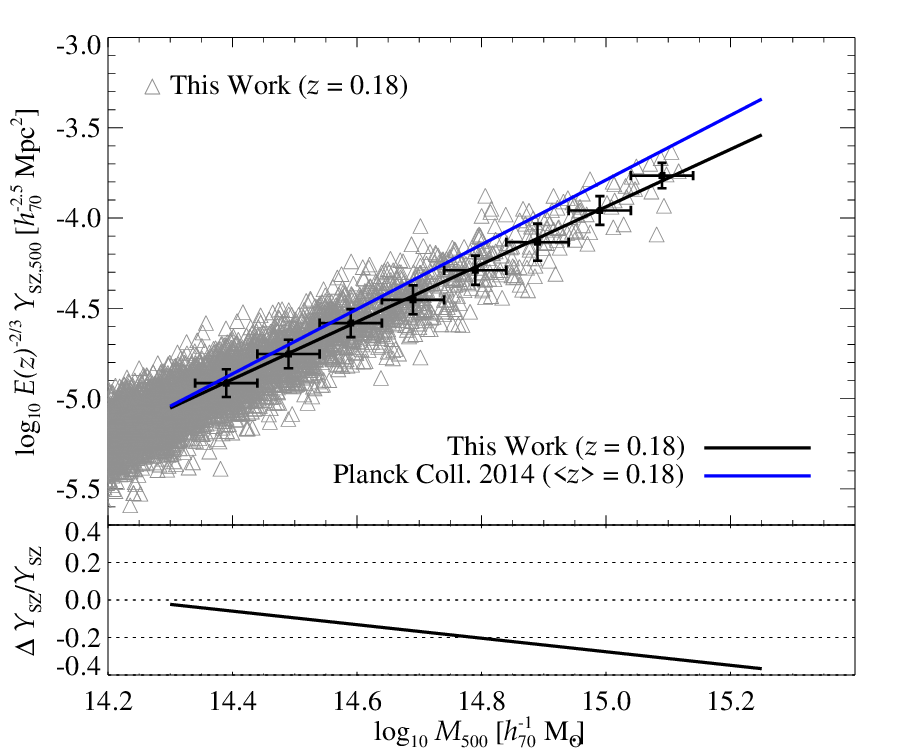}
\includegraphics[width=0.48\textwidth]{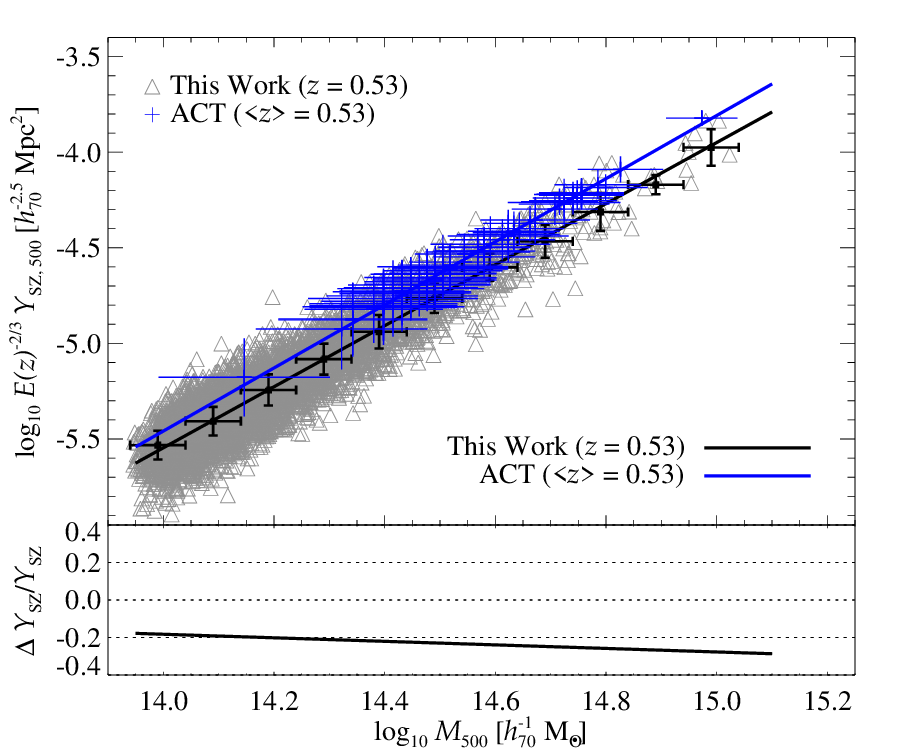}
\caption{Comparison of our MultiDark SZ scaling relations (grey triangles)
    with data. The black crosses represent the $Y_{\rmn{SZ}}$ median values of
    our sample for a given mass bin (indicated by horizontal error bars), and
    the vertical error bars represent the standard deviation within a bin.  The
    bottom panels show the relative difference to the observational scaling
    relations.  \emph{Left.}  We compare our $Y_{\rmn{SZ}}-M$ relation at
    $z=0.18$ with the {\em Planck} data \citep[blue,][]{2013arXiv1303.5080P},
    which have a median redshift of about $0.18$. \emph{Right.} We compare our
    $Y_{\rmn{SZ}}-M$ relation at $z=0.53$ with the ACT sample
    \protect\citep[blue,][]{2013JCAP...07..008H} assuming a self-similar
    redshift evolution.}
\label{fig:SZ_M}
\end{figure*}

In the left panel of Fig.~\ref{fig:SZ_M}, we compare the $Y_{\rmn{SZ}}-M_{500}$
relation of our model (following \citealp{2012ApJ...758...74B}, equation~3) with
the observed scaling relation by the \cite{2013arXiv1303.5080P}. We use the
  \emph{Planck} COSMO sample, which is Malmquist-bias corrected and has a median
  redshift of $0.18$, and compare this to our $z=0.18$ relation. Our model
reproduces the normalization of the scaling relation remarkably well, except for
the high-mass end where our simulations have a weaker constraining power due to
the smaller box size in comparison to the survey volume of {\em Planck}.
  However, the slope of the MultiDark $Y_{\rmn{SZ}}-M_{500}$ relation ($1.59\pm0.01$) is
  shallower in comparison to the self-similar slope ($5/3$) as well as the
  observed slope of the \emph{Planck} COSMO sample. We can analytically
  understand the cluster-mass scaling of our model by considering its
  mass-dependent quantities, namely $Y_{\rmn{SZ}}\propto M_{500} f_{\rmn{gas}}
  T \propto M_{500}^{1.595}$. In particular, we can trace back the
  shallower slope to the adopted temperature scaling of our model
  ($T\approx T_{\rmn{ci}}\propto M_{500}^{0.46}$), which includes the cluster core region
  and yields a shallower mass scaling in comparison to the self-similar
  expectation ($T\propto M_{500}^{2/3}$). The temperature scaling is only
  partially countered by the scaling of the gas mass fraction with cluster mass,
  $f_{\rmn{gas}}\propto M_{500}^{0.135}$. To improve upon this, we would have to
  account for a slightly steeper temperature-mass scaling relation (e.g., using a
  centrally excised temperature scaling, see \citealt{2010MNRAS.406.1773M}) in
  the outer cluster parts that are of relevance for the SZ flux, as explained in
  Sect.~\ref{sec:Xscaling}. We find a scatter of $\sigma_{yx} \approx 0.11$
which compares well with the \emph{Planck} result of $\sigma_{yx} \approx
0.08$. In Table~\ref{tab:YSZfits}, we report our SZ scaling relations for
different redshifts.

\begin{table} 
\begin{center}
\caption{$Y_{\rmn{SZ}, 500}-M_{500}$ scaling relations.}
\medskip
\begin{tabular}{cccc}
\hline
\phantom{\Big|}
redshift $z$ & $A$ & $B$ & $\sigma_{yx}$ \\
\hline\\[-0.5em]
 0      & $-27.93\pm0.07$ & $1.60\pm0.01$ & 0.109\\
 0.1   & $-27.79\pm0.07$ & $1.59\pm0.01$ & 0.109\\
 0.18 & $-27.82\pm0.08$ & $1.59\pm0.01$ & 0.109\\
 0.2   & $-27.76\pm0.08$ & $1.59\pm0.01$ & 0.109\\ 
 0.4   & $-27.79\pm0.10$ & $1.59\pm0.01$ & 0.108\\
 0.53 & $-27.90\pm0.12$ & $1.60\pm0.01$ & 0.109\\
 0.61 & $-27.86\pm0.13$ & $1.59\pm0.01$ & 0.109\\ 
 0.78 & $-27.62\pm0.18$ & $1.58\pm0.01$ & 0.109\\ 
 1      & $-27.38\pm0.26$ & $1.58\pm0.02$ & 0.109\\[0.5em] 
\hline
\end{tabular}
\label{tab:YSZfits}
\end{center}
\footnotesize{Note. Scaling relations are reported in the form of $\log_{10}~(E(z)^{-2/3}~Y_{\rmn{SZ},500}~/~h_{70}^{-2.5}~\rmn{Mpc}^{2})=A+B~\log_{10}~(M_{500}~/~h_{70}^{-1}~\rmn{M_{\odot}})$. The relation scatter $\sigma_{yx}$ is also shown.}
\end{table}

  In the right panel of Fig.~\ref{fig:SZ_M}, we also compare our scaling relation
  to the recent Atacama Cosmology Telescope (ACT) sample
  \citep{2013JCAP...07..008H}, which has a median redshift of $z=0.53$.  
  We conclude that overall there is reasonable
  agreement between our phenomenological model and the SZ data of the
  \emph{Planck} and ACT collaborations.

\section{Multi-frequency Mock Catalogs}
\label{sec:6}

As a final product of this work, we construct cosmologically complete mock
  catalogs of galaxy clusters at different redshifts (see Table~\ref{tab:z}) for
  $M_{200}\geq1\times10^{14}$~$h^{-1}\rmn{M}_{\odot}
  \approx1.4\times10^{14}$~$h_{70}^{-1}$~M$_{\odot}$.  We make these catalogs
  publicly available through the MultiDark database (\textit{www.multidark.org})
  where we will also post possible updates.

  Our catalogs contain all the information regarding the DM properties of each
  cluster as given by the corresponding original MultiDark BDM halo
  catalogs. Additionally, we include the following information:

\begin{itemize}
\item $M_{500}$ and $R_{500}$ calculated from $M_{200}$ and $R_{200}$ of the BDM catalogs with the \cite{2003ApJ...584..702H} method,
\item the ICM X-ray temperature assigned via the $T_{\rmn{ci}}-M_{500}$ relation by \cite{2010MNRAS.406.1773M},
\item a flag identifying the cluster as NCCC or CCC, as described in Section~\ref{sec:3},
\item the central gas density $\rho_{\rmn{gas},0}$ with which the full ICM radial profile can be calculated as in Section~\ref{sec:3},
\item the bolometric X-ray thermal bremsstrahlung luminosity $L_{\rmn{X, bol}}$
  within $R_{500}$, and
\item the $Y_{\rmn{X}}$ and $Y_{\rmn{SZ}}$ parameters within $R_{500}$.
\end{itemize}

In Paper~II, we present a model that allows us to compute the possible radio and
gamma-ray emission for each cluster in our mock catalog, as result of the
cosmic-ray (CR) proton-proton interactions with the ICM. We refer to that paper
for all the details. However, we want to point out that our mock catalogs also
include:

\begin{itemize}
\item a flag identifying whether a given object is a radio-loud or a radio-quiet cluster, 
\item the CR-to-thermal pressure averaged over the cluster volume within $R_{500}$,
\item the radio synchrotron luminosity of secondary electrons that are produced
  in hadronic CRp-p interactions, within $R_{500}$, at 120~MHz and 1.4~GHz,
\item the gamma-ray luminosity, within $R_{500}$, above 100~MeV and 100~GeV.
\end{itemize}

We note that our mock catalogs are not appropriate for all purposes because the
density profiles that we adopt in our model represent the average cluster
population of cool-core and non cool-core systems (while we account for scatter
in the gas mass fractions, i.e., the normalization of the profiles). As such,
our mocks are very useful as a baseline for comparisons to new hydrodynamical
simulations and to new observational samples, as well as for non-thermal
modeling of clusters. On the contrary, in the current form they are not
appropriate to perform detailed cosmology studies. Careful modeling of the
response function of the considered instrument, in combination with light-cone
mock catalogs, would be required in this case. Nevertheless, for volume-limited
cluster samples, our mock catalogs will certainly serve as valuable tools.

\section{Discussion and Conclusions}
\label{sec:7}

  We build a complete cosmological sample of galaxy clusters from the
  MultiDark $N$-body simulation with redshifts ranging from $z = 0$ to 1 and
  construct a \emph{phenomenological} model for the ICM. This is characterized
  by X-ray-inferred (CCC and NCCC) gas profiles (taken from the REXCESS sample)
  and a cluster mass-dependent gas fraction. Note that our model is calibrated
  with X-ray observables within $R_{500}$. More work would be needed to
  extrapolate the model to larger cluster radii that are characterized by an
  increased complexity, which manifests itself in a larger kinetic-to-thermal
  pressure support, an increased level of density clumping, as well as cluster
  asphericity and ellipticity. In our model, we assign a (cluster
  mass-dependent) gas density profile to each DM halo in our sample and sort it
  into NCCC/CCC populations according to a dynamical disturbance parameter
  that is calculated from the DM distribution. Applying the model to our sample
  of cluster halos, we obtain volume-limited mock catalogs for the thermal (SZ
  and X-ray) cluster emission. Our mock catalogs match the observed X-ray
  luminosity function as well as the $L_{\rmn{X, bol}}-M_{500}$,
  $Y_{\rmn{X}}-M_{500}$, and $Y_{\rmn{SZ}}-M_{500}$ relations well.

  However, there are some deviations among the different observational scaling
  relations and our model, which may either point to observational sample
  selection effects that are not accounted for or missing complexity of our
  model. This model was specifically constructed to provide a reliable
  description of the gas density, which necessarily implies a great match of the
  resulting mock X-ray observables to data (scaling relations and luminosity
  functions), as well as to the low-redshift and massive cluster sample probed by
  {\em Planck}. However, we note that our model has a slightly shallower slope
  of the $Y_{\rmn{SZ}}-M_{500}$ relation, which implies that the X-ray inferred
  deviations of self-similarity on small scales become weaker on the larger
  scales probed through the SZ effect. 

  In the companion Paper~II, we additionally present a model for the radio and
  gamma-ray emission as a result of hadronic CR interactions with the ICM.  We
  provide cosmologically complete multi-frequency mock catalogs for the
  \mbox{(non-)thermal} cluster emission at different redshifts and make these
  catalogs publicly and freely available on-line through the MultiDark database
  (\textit{www.multidark.org}). Our mock catalogs are based on density profiles
  that represent the {\em average} cluster population of cool-core and non
  cool-core systems (while we account for scatter in the gas mass fractions,
  i.e., the normalization of the profiles). As a result, our model does not
  fully exploit the true variance across the average profile shapes which may
  result in a limited variance of thermal cluster observables. Nevertheless, the
  mock catalogs will allow quantitative comparisons to future hydrodynamical
  simulations and forecasts for future surveys.  We emphasize, however, that the
  catalogs are not suited for detailed cosmological studies in their (present)
  published form, which do not address any survey response functions nor provide
  light-cone mock catalogs. However, future surveys should be able to test the
  underlying assumptions of our modeling approach, which is based on X-ray
  observations of clusters at low redshift ($z<0.2$) and adopts self-similar
  redshift evolution for extrapolating to higher redshifts. We hope that the
  community can make valuable use of these catalogs in synergy with the future
  radio, X-ray and gamma-ray data.

\section*{Acknowledgments}
We thank the anonymous referee for the useful comments.
We thank Nick Battaglia, Harald Ebeling, Stefan Gottl{\"o}berg, Anatoly Klypin,
Andrey Kravtsov, Adam Mantz, Frazer Pearce, Jos\'e Alberto Rubi\~no, and Gustavo
Yepes, for the useful discussions. We also thank the MultiDark database people,
in particular Adrian Partl and Kristin Riebe.  F.Z.{\ }acknowledges the CSIC
financial support as a JAE-Predoc grant of the program ``Junta para la
Ampliaci\'on de Estudios'' co-financed by the FSE.  F.Z.{\ }acknowledges the
hospitality of the Leiden Observatory during his stay.  F.Z.{\ }and F.P.{\ }
thank the support of the Spanish MICINN's Consolider-Ingenio 2010 Programme
under grant MultiDark CSD2009-00064, AYA10-21231. C.P.{\ }gratefully acknowledges financial
support of the Klaus Tschira Foundation. The MultiDark Database used in this
paper and the web application providing online access to it were constructed as
part of the activities of the German Astrophysical Virtual Observatory as result
of a collaboration between the Leibniz-Institute for Astrophysics Potsdam (AIP)
and the Spanish MultiDark Consolider Project CSD2009-00064, AYA10-21231. 
The Bolshoi and MultiDark simulations were run on the NASA's Pleiades supercomputer 
at the NASA Ames Research Center.

\bibliographystyle{mnras}
\bibliography{bib_file}

\label{lastpage}

\end{document}